\documentclass[doublespacing]{elsart}
\usepackage{graphicx}

\usepackage{amssymb}

\begin{document}

\begin{frontmatter}



\title{Link between the diversity, heterogeneity and kinetic properties
of amorphous ice structures}


\author{M.~M.~Koza, T.~Hansen, R.~P.~May, H.~Schober }

\address{Institut Laue--Langevin, 6 rue Jules Horowitz,
F--38042 Grenoble Cedex 9, France.}

\begin{abstract}
Based on neutron wide--angle diffraction and small--angle neutron scattering
experiments, we show that there is a correlation between the preparational conditions
of amorphous ice structures, their microscopic structural properties, the extent of
heterogeneities on a mesoscopic spatial scale and the transformation kinetics.  
There are only two modifications that can be identified as homogeneous disordered
structures, namely the very high--density vHDA
and the low--density amorphous LDA ice.
Structures showing an intermediate static structure factor with respect to vHDA and LDA
are heterogeneous phases.
This holds independently from their preparation procedure,
i.e. either obtained by pressure amorphisation of ice I$_{\rm h}$ or by heating
of vHDA.
The degree of heterogeneity can be progressively suppressed when higher pressures
and temperatures are applied for the sample preparation.
In accordance with the suppressed heterogeneity the maximum of the static structure
factor displays a pronounced narrowing of the first strong peak, shifting towards
higher $Q$--numbers.
Moreover, the less heterogeneous the obtained structures are the slower is the
transformation kinetics from the high--density modifications into LDA.
The well known high--density amorphous structure HDA
does not constitute any particular state of the amorphous water network.
It is formed due to the preparational procedure working in liquid
nitrogen as thermal bath, i.e. at about 77~K.
\end{abstract}

\end{frontmatter}

\section{Introduction}
\label{introduction}
Amorphous polymorphism is a property of solid water well
established by experiments.
A metastable high--density amorphous structure
(HDA, $\rho\approx38$~molec./nm$^3$) can be formed by pressure
amorphisation of hexagonal ice~I$_{\rm h}$
at 77~K \cite{Mishima-Nature-1984}.
HDA transforms upon heating at ambient pressure to a low--density 
amorphous modification (LDA, $\rho\approx31$~molec./nm$^3$).
If higher temperatures are applied during the pressure
amorphisation of ice~I$_{\rm h}$ or if the formed HDA is
subject to a heat treatment at high pressures
amorphous structures of higher densities are formed
\cite{Mishima-Nature-1996,Loerting-PCCP-2001}.
A very high--density structure (vHDA, $\rho\approx 41$~molec./nm$^3$
has been conjectured to be a third 'state' of amorphous ice 
\cite{Loerting-PCCP-2001}.
The number of next neighbours has been proposed as a discrimination
criterion between the amorphous 'states' LDA, HDA and vHDA
from neutron diffraction experiments
\cite{Finney-PRL-2002}.

The assignment of states to the amorphous ice structures becomes
less evident, when transformation properties between the structures
are monitored.
For example, it has been shown that an HDA sample transforming
upon heating into LDA displays a continuum of intermediate static
structure factors \cite{Schober-PB-1998,Tulk-Science-2002,Koza-JPCM-2003}
and dymanic properties \cite{Schober-PB-1998,Koza-PRL-2005}.
The transformation of vHDA into LDA passes through a structure
displaying a structure factor apparently comparable to
the one of HDA which, however, does not feature any distinguished
properties among other intermediate structures \cite{Koza-PRL-2005}.

As far as diffraction experiments are concerned, the evolution
of a small--angle neutron scattering (SANS) signal represents
rather a counter evidence against the scenario of a multiple
of amorphous states.
When HDA and vHDA transform into LDA a transient excess in
the SANS signal marks the intermediate transformation
stages as heterogeneous structures \cite{Schober-PB-1998,Koza-PRL-2005}.
Fingerprints of pronounced heterogeneity can be already observed
in HDA samples and only vHDA and LDA proved to be homogeneous
and, hence, good candidates for distinguished states
\cite{Koza-PRL-2005,Schober-PRL-2000}.
The properties of the wide-angle diffraction (WAD) signal
indicate that spatial correlations are notedly confined
in the intermediate structures, and the evolution
of the spatial confinement compares well with the
time dependence of the SANS intensity.

In this paper we intensify our studies of the significance
of preparation conditions on the WAD and SANS properties.
We show that the static structure factor $I(Q)$ of the amorphous
ice structures shows a systematic dependence in the sense
that the first structure peak shifts towards higher
momentum numbers and becomes narrower if higher pressures
and/or temperatures are applied.
In accordance with the narrowing of the peak,
which could be interpreted as an increment of
correlations in the amorphous structure, the excess intensity
in the SANS signal is reduced, marking the structure as less
heterogeneous.

{\it In situ} observation of structural changes $I(t,T)$
reveals a progressive slowing down of transformation kinetics
for samples prepared at higher temperatures and pressures. 
Regardless of the applied preparation conditions
all samples show comparable structural properties when
transforming into LDA.
The average size of mesoscopic domains leading to the structure
of strongest heterogeneity (SSH) is estimated to about 12~\AA .

\section{Experimental}
\label{experimental}

Six samples (D$_2$O, purity of 99.8\%) were prepared by pressure
amorphisation of ice~I$_{\rm h}$ at $T\approx 77$~K.
The maximum pressure of 18~kbars was applied for 10~min.
before a lower pressure $p_{\rm anneal}$ was established
at which the samples were annealed at $T_{\rm anneal}$
for 30~min. each.
After the annealing process all samples were cooled back
to 77~K and recovered from the pressure device.
An overview of the sample preparation conditions
is given in table~\ref{tab_01}.
In addition, a seventh sample was obtained directly by pressure amorphisation
at $T\approx 120$~K with a maximum pressure of 15~kbar.
It had been equally kept for 30~min. at these conditions before
it was recovered at 77~K.
The temperature and pressure stability was $\pm 2$~K
and $\pm 0.25$~kbar, respectively.

Each of the preparation runs resulted in about 3~ml of sample
substance.
Each sample was separated into 5--6 portions and used for different
experiments, either at different instruments or at the same 
instrument but at different experimental conditions.
The experiments were performed at the wide--angle diffractometer
D20 and the small--angle diffractometer D22 at the
Institut Laue--Langevin, Grenoble, France.
The incident neutron wavelengths used were 2.4~\AA\ and 6.0~\AA ,
respectively. 
In all measurements standard cryostats were utilised with
a He--atmosphere of 200~mbars.
The {\it in situ} experiments and the data treatment were
performed in full analogy to the measurements described
in detail in references \cite{Koza-JPCM-2003,Koza-PRL-2005}.
Two temperatures were chosen to follow the structural
transformation of the samples {\it in situ}, namely 108~K and 113~K.
All measurements were finished with the LDA structure
annealed for 30~min. at 130~K.

To compare all results on a relative scale all WAD data
were normalized with respect to the coherent scattering power of LDA.
In addition, to compare with SANS data the results
have been shifted to give zero scattering towards small
scattering angles.

\section{Results and Discussion}
\label{results}
\begin{figure}
\includegraphics[angle=0,width=140mm]{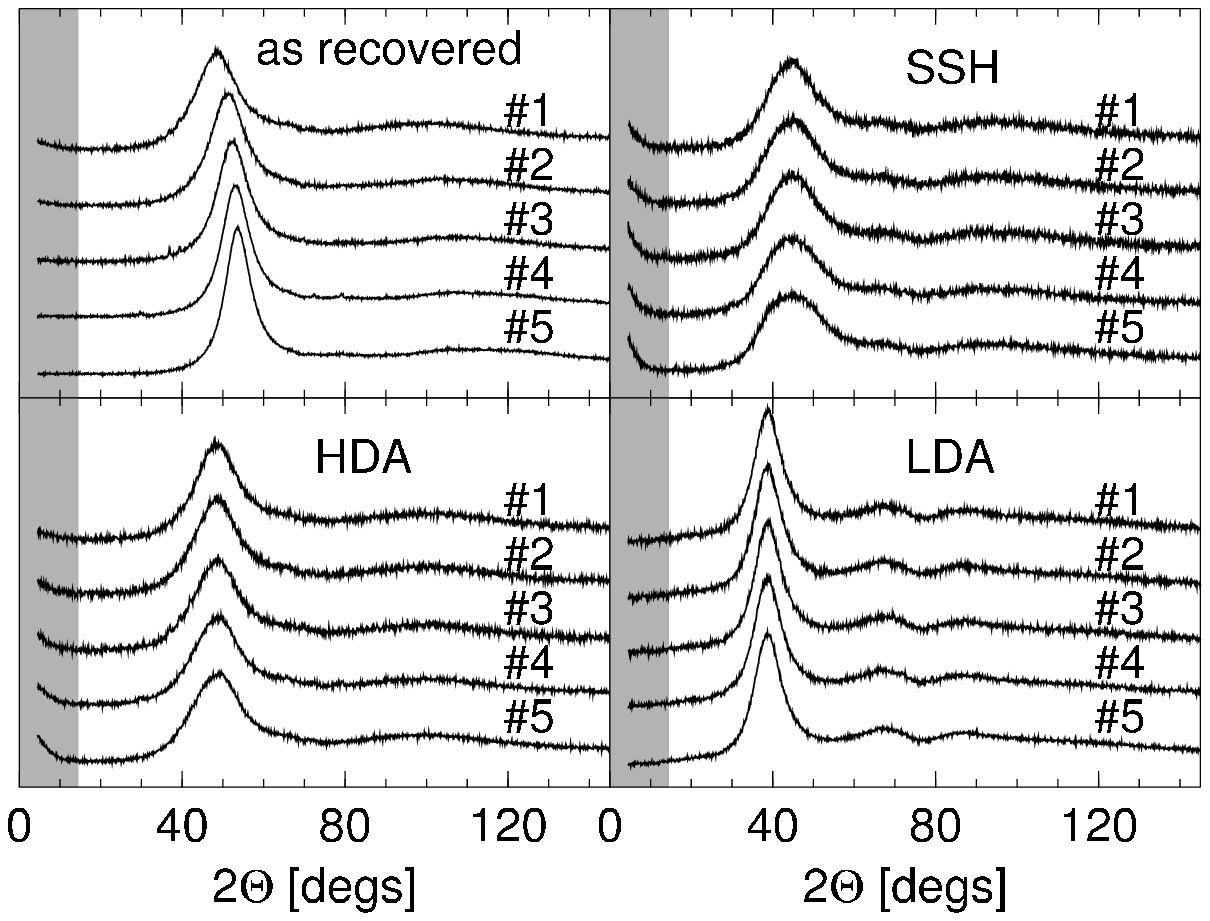}
\caption{Static structure factor of samples \#1, \#2, \#3, \#4 and
\#5 as they are recovered, obtained after a certain time 
in a state comparable with HDA and SSH, and after they
were annealed into the LDA structure.
The respective data sets have been shifted for clarity.}
\label{fig_01}
\end{figure}

In Fig.~\ref{fig_01} we present exemplary static structure factors
of samples \#1--\#5.
Depicted are results from the 'as recovered' states, from stages of
transformation comparable to the HDA structure, from the structures
of strongest heterogeneity (SSH), and from LDA.
It is obvious that beyond the differences of the 'as recovered'
structures there is a common behaviour among all samples
when transforming into LDA.
Besides the well reported evolution of the predominant maximum
that shifts progressively towards lower scattering angles \cite{Koza-JPCM-2003},
and takes on temporarily the lowest intensity and largest
width in the middle of the transformation,  
there is a transient excess intensity in the small--angle
part of the diffractograms.
The regime of the changing SANS signal is highlighted by
the grey area.

There are two important points to be taken note of
in the properties of the SANS signal.
First of all, its intensity reaches a maximum right in the
middle of the transformation and marks this stage
as a structure of strongest heterogeneity (SSH).
This SSH matches well with the stage at which the
structure factor maximum is characterised by the largest
width and lowest intensity, as it is reported in
\cite{Koza-PRL-2005}.
These properties are independent of the sample preparation
conditions.

\begin{figure}
\includegraphics[angle=0,width=140mm]{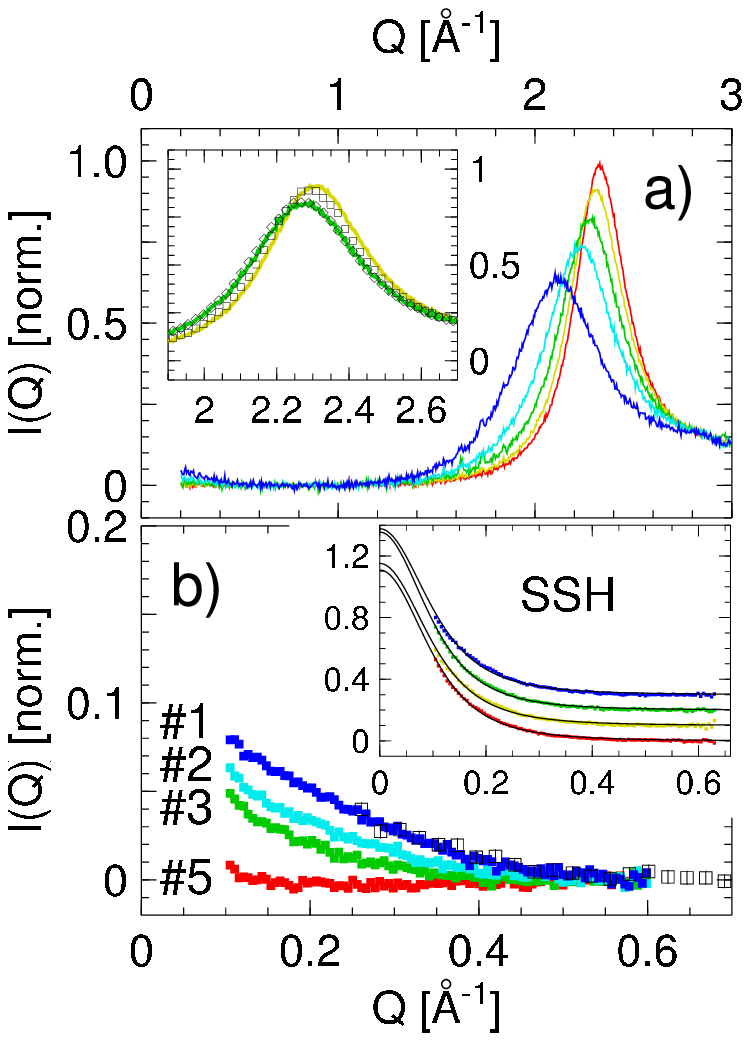}
\caption{Figure a: WAD signal of the 'as recovered' samples.
The samples correspond from left to right of the maximum position
\#1 (blue), \#2 (cyan), \#3 (green), \#4 (yellow)
and \#5 (red).
The inset indicates the resemblance of samples \#6 (open squares) and
\#7 (open diamonds) with \#4 and \#3, respectively.
Figure b: SANS signal of the 'as recovered' samples indicated
in the figure. Open squares show for comparison data
of sample \#1 from the D20 measurement.
The inset reports the SSH signal and the DBM fit results, explained
in the text.
The size of data points reflects their statistical error.}
\label{fig_02}
\end{figure}

The second important point concerns the 'as recovered'
structures, i.e. properties that depend on sample preparation
details.
As can be seen from the 'as recovered' data in Fig.~\ref{fig_01} 
only samples \#4 and \#5 display below 30~degs. a flat
scattering characteristic, without any excess intensity.
In samples \#1--\#3 an upturn of the signal towards
low scattering angles indicates the presence of heterogeneities.
To substantiate this point, Fig.~\ref{fig_02}~a reports
the WAD data in comparison with the SANS results
in Fig.~\ref{fig_02}~b taken on different portions of the
same samples.
Please note, that the SANS response of sample \#4 matches
the signal of \#5 and has been therefore suppressed
in Fig.~\ref{fig_02}~b.
There is an obvious and unequivocal correlation between
the excess intensity in the SANS data and the position
and width of the maximum in the WAD data, respectively.
We may conclude that the higher the temperature and/or
the higher the pressure during the sample preparation
is, the less is the SANS excess intensity, the narrower
is the static structure maximum and the higher is its
position in the diffractogram.

In other words, only when extreme conditions for the
formation of the amorphous ice structures are applied
homogeneous samples can be obtained.
HDA, as sample \#1, and samples \#2, and \#3 are
structurally heterogeneous.
To stress the relation between the signal at low--scattering angles
and the width of the static structure factor peak, table~\ref{tab_02}
reports data on the intensity added up as
$I_{\rm sas} = \sum_{0.2\AA^{-1}}^{0.6\AA^{-1}} I(Q)$ and
the full width at half maximum (FWHM) obtained from
a Lorentzian fit to the peak.
Not only the  $I_{\rm sas}$ and FWHM of samples
\#1--\#5 follow a common trend, but also
samples \#6 and \#7 fit into this picture.
The inset in Fig.~\ref{fig_02}~a reports the peak in
$I(Q)$ of \#6 and \#7 indicating the ressemblance with
\#2 and \#3.
As it is demonstrated by the properties of sample \#7,
pressure amorphisation at higher temperatures
than 77~K leads to structures that are less heterogeneous
than HDA, however, still displaying an enhanced SANS signal
and a clearly broadened peak in the static structure factor.

It is important to note that a recent publication of static
structure factors of amorphous ice modifications which have been
formed by compression at $T = 125$~K reports an ascending signal
towards smaller scattering angles, i.e., smaller $Q$--numbers
for the intermediate structure only \cite{Loerting-PRL-2006}.
The intermediate structure is denoted as {\it sample B} in the publication.
The recorded small--angle signal is reminiscent of published
structure factors from X-ray scattering experiments \cite{Schober-PRL-2000}.
It is therefore reasonable to conclude that {\it sample B}
is equally of heterogeneous character.

In order to extract a measure of the extent of the heterogeneous domains
we have approximated the SANS signal of the samples in their
SSH with the Debye--Bueche model (DBM) \cite{Debye-JAP-1949}.
The DBM describes a mixture of two statistically
distributed phases leading to the SANS formfactor
$I(Q) = I_{\circ}/(1+(Q\cdot\gamma)^2)^2$.
The correlation length $\gamma$ takes on the simple
relation $2\gamma = D$ with the average domain size
$D$ when the occupation numbers of both phases is 50~\%,
and, thus, the sample in the SSH stage.
Examplary fits are depicted in the inset of Fig.~\ref{fig_02}~b.
The data sets have been shifted for clarity. 
Throughout all the monitored SSH we arrive at a $D$
of 11--13~\AA .
Within the DBM this small number indicates that structural
correlations are confined in the SSH to a space smaller
than they are in the homogeneous states, i.e. vHDA and LDA.
This result is in full agreement with the extended spatial
correlations of vHDA and LDA in comparison to SSH and HDA
\cite{Koza-PRL-2005,Guthrie-CPL-2004}.
The absolute value of $D$ corresponds amazingly well
with the cutoff distance above which no oscillations
are observed in real space distribution functions 
of SSH \cite{Koza-PRL-2005}.
As a clear consequence, sample properties that are essentially
determined by the correlation length are altered progressively by
the 'heterogeneous' confinement.
For example, the WAD signal of the intermediate structures
will not be characterisable by a superposition of the
homogeneous structures vHDA and LDA.
Hence, the behaviour of the static structure factors
does not allow to draw any conclusion upon the nature
of the phase transitions between the amorphous
ice modifications.

\begin{figure}
\includegraphics[angle=0,width=140mm]{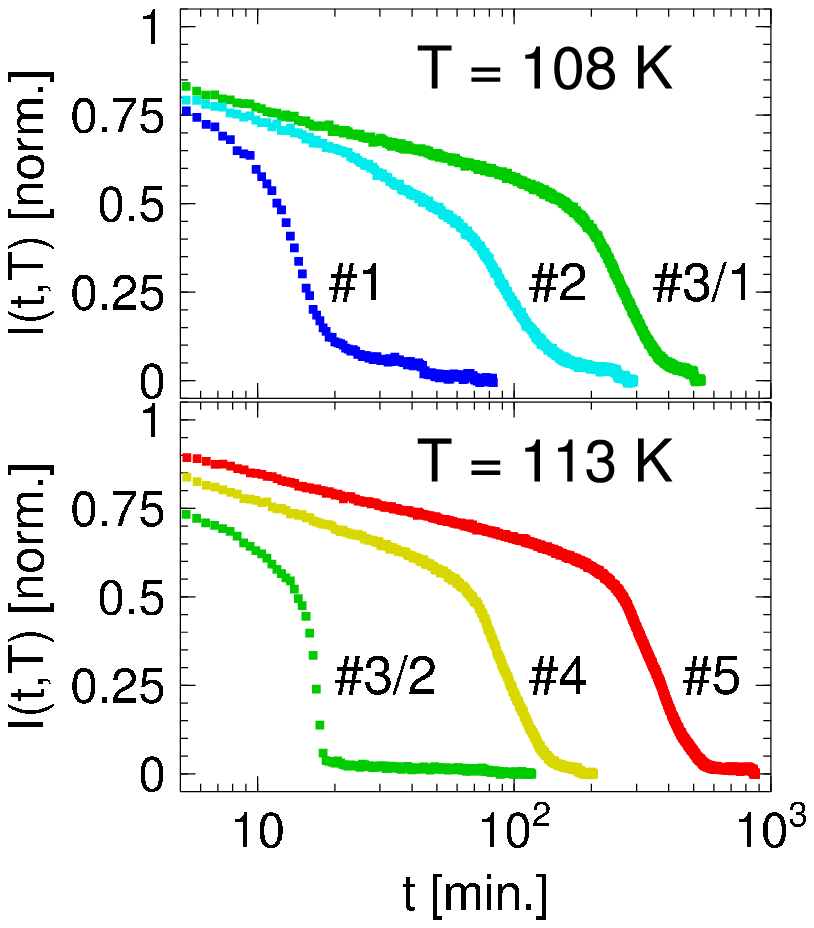}
\caption{Transformation kinetics of samples \#1--\#5 into the LDA 
modification measured at two different temperatures.
Two portions of sample \#3 are applied as reference at
both temperatures.}
\label{fig_03}
\end{figure}

Finally we address the transformation kinetics of the
different amorphous ice structures into LDA.
We visualize the kinetics in Fig.~\ref{fig_03} by 
the function $I(t,T)$, which has been discussed in detail
in \cite{Koza-JPCM-2003}.
$I(t,T)$ takes on the value 1 for the initial
transformation stages, i.e. the 'as recovered'
structures.
The final LDA states are characterised by 0.
Due to the strong variation in kinetic properties of
the samples we had to follow the transformations
at two different temperatures, whereby two
portions of sample~\#3 served as reference
at both $T$.
They are marked as \#3/1 and \#3/2.
Please note that small steps at the end of each data
set are due to the annealing of the samples into
the LDA structure at $T=130$~K.

We may easily conclude from the systematic shift of
the characteristic transformation times that
the less heterogeneous a sample appears to be,
the slower is its kinetics when transforming
into LDA.
This dependence can be intuitively conjectured,
on the one hand, from the Adam--Gibbs theory,
which predicts that relaxation times are slowed
down for systems with lower configurational entropy
\cite{Adam-JCP-1965}.
The excess SANS intensity of, at least, samples \#1 (HDA),
\#2 and \#3 indicates an entropy elevated above that
of samples \#4 and \#5.
However, so far it is not clarified whether we deal
in the case of the amorphous ice structures with glassy
modifications of liquid water to which the Adam--Gibbs
approach might be applicable.
There is indeed a number of computer studies questioning
the link between the amorphous polymorphism of
ice with the liquid state of water
\cite{Shpakov-PRL-2002,Giovambattista-PRL-2003}.

On the other hand, if we consistently interprete the
amorphous structures in a two--phase scenario,
which finds some justification by their heterogeneous
character, different initial mixing ratios of two
homogeneous phases might equally lead to distinctly
different transformation kinetics.
This should hold for a transformation comprising strong volume
changes as it is the case here \cite{Tanaka-EPL-2000}.

\section{Conclusions}
\label{conclusions}

We have shown that only the application of
extreme conditions in terms of pressure and temperature
for the preparation of amorphous ice structures lead to
samples which are homogeneous.
There is a number of indicators for the grade of sample
homogeneity accessible in diffraction experiments.
These are, for example, the excess scattering intensity
in the small--angle regime, the position and width of
the strong maximum in the static structure factor,
and the characteristic transformation times in {\it in situ}
studies.

The properties of the SANS intensity offer the opportunity
of discriminating between distinguished 'states' of a substance,
which are expected to be homogeneous.
Only the structures LDA and the sample \#5, fullfill this
criteria.
However, please note that the best homogeneous structure,
i.e. the best candidate for a 'state', is not known.
The well known modification HDA is a heterogeneous
structure and, hence, not a 'state'.
Its apparent distinctiveness is only due to the convenient
application of liquid nitrogen for thermalisation of
the sample during its preparation.

There is a clear correlation between the grade of
heterogeneity of a structure and the properties of
its wide--angle diffraction response.
As a consequence, the interpretation of the intermediate
structures comprising also HDA in real space is not a
straightforward procedure.
The excess SANS signal should, however, help to discern
between different models trying to account for their properties.

The problem of characterising the static properties of the
intermediate structures in real space can be well compared
with the problem of finding the origin of the non--exponential 
time response of relaxiations in glasses.
There is still no general perception on whether the non--exponential
ensemble--averaged time response, is due to a homogeneous or heterogeneous
distribution of energy barriers determining the relaxation processes
\cite{Giovambattista-JPCB-2004}.
It is obvious that in the heterogeneous case there is no general
way of characterising the system in a unified manner
since each sub--ensemble experiences a different relaxation
scenario.
In fact, the present and previous findings on the small--angle signal
\cite{Schober-PB-1998,Koza-PRL-2005,Schober-PRL-2000}
demonstrate that the heterogeneous scenario is applicable in the case
of the intermediate amorphous ice structures.
Thus, real space distribution functions obtained as 
Fourier--transforms of the ensemble--averaged static structure factor
do not comprise the complexity of the real space structure
correctly.
The use of only three radial distribution functions
to describe the correlations between Oxygens, Hydrogens
and cross--terms might be only justified for the 
LDA and vHDA as homogeneous structures, hence,
if the ensemble--averaged static structure factor
resembles the characteristics of each of the sub--ensembles.

Finally, results which can be extracted in absolute units
from the presented data must not be mistaken as the consequence
of the annealing pressure and temperature only, which were
applied during the sample formation.
Experimental parameters like the compression rate, the annealing
time, the decompression procedure, purity grade of the sample
substance, etc. may have a well noticeable influence on the
sample properties.
The experiments reported here were done in a consistent way
within one week only including sample preparation and the entire
sets of measurements.
They should give therefore an accurate qualitative description
of the amorphous ice sample properties.


\begin{table}
\caption{Sample designation according to the nominal annealing temperature
$T_{anneal}$ and annealing pressure $p_{anneal}$.
Please note that sample \#7 was first heated to 120~K before
having been compressed to 15~kbars.
All other samples were first compressed at 77~K
before having applied $T_{anneal}$.}
\begin{center}
\begin{tabular}{|c||c|c|c|c|c|c|c|}\hline
 & \#1 & \#2 & \#3 & \#4 & \#5 & \#6 & \#7 \\ \hline\hline
$T_{anneal}$/K & 77 & 100 & 120 & 140 & 150 & 140 & 120 \\ \hline
$p_{anneal}$/kbar & 15 & 15 & 15 & 15 & 15 & 12 & 15 \\ \hline
\end{tabular}
\end{center}
\label{tab_01}
\end{table}
\begin{table}
\caption{Summed intensity in the small--angle regime $I_{\rm sas}$
from the WAD data on the 'as recovered' samples, FWHM and position
P of the strong maximum from Lorentzian fits to the WAD signal.}
\begin{center}
\begin{tabular}{|c||c|c|c|c|c|c|c|}\hline
 & \#1 & \#2 & \#3 & \#4 & \#5 & \#6 & \#7 \\ \hline\hline
 $I_{\rm sas}$ & 1.51(8) & 0.64(8) & 0.35(6) & 0.05(6) & 0.08(5) & -0.07(9) & 0.51(5) \\ \hline
FWHM/\AA $^{-1}$ & 0.558 & 0.452 & 0.406 &  0.355 & 0.326 & 0.361 & 0.400 \\ \hline
P/\AA $^{-1}$ & 2.120 & 2.232 & 2.278 & 2.316 & 2.337 & 2.301 & 2.271 \\ \hline
\end{tabular}
\end{center}
\label{tab_02}
\end{table}
%




\end{document}